\documentclass[epj,nopacs]{svjour}
\usepackage{graphics}
\usepackage{epsfig}
 
\def\Journal#1#2#3#4{{#1} {\bf #2}, #3 (#4)}

\def\PRD{{\em Phys. Rev.} D}

\def\EPJ{\em Eur. Phys. J.}
\def\JPG{\em J. Phys. G: Nucl. Part. Phys.}

\begin{document}
 
\title{Search for Extra Dimensions at LHC}
\author{Laurent Vacavant, for the ATLAS and CMS Collaborations}
\institute{Lawrence Berkeley National Laboratory, Physics Division, Berkeley CA 94720, USA}
\date{Received: \today}
\abstract{
Some of the studies performed by the ATLAS and CMS collaborations to establish 
the future sensitivity of the experiments to extra dimension signals are reviewed.
The discrimination of those signals from other new physics signals and 
the extraction of the underlying parameters of the extra dimension models are 
discussed.
\PACS{
      {}{}   \and
      {}{}
     } 
} 
\maketitle
\section{Introduction}
\label{intro}

Models with extra dimensions (ED) (see~\cite{intro}) 
are very attractive extensions of the Standard Model (SM), 
in particular with respect to the 
hierarchy problem.
While ED have so far escaped detection~\cite{limits}, 
they could manifest themselves at LHC via a rich and varied 
phenomenology.
 
This review focuses on the three main classes of ED models 
with prediction at the TeV scale.
The aim of the studies was two-fold: establishing the sensitivity 
to ED signals using detector simulation and various 
physics backgrounds, as well as assessing whether enough information 
could be extracted in order to distinguish ED signatures from 
other new physics signals.

The studies are based on fast simulation tools which describe 
accurately the expected detector performance. The relevant aspects of the 
simulation have been validated~\cite{tdr} in full simulation and with test-beam data 
whenever possible.
Except when stated otherwise, all the results and plots presented here are 
for an integrated luminosity of 100 fb$^{-1}$ collected by one of the experiments 
(i.e. one year at the nominal luminosity 
of LHC, 10$^{34}$ cm$^{-2}$s$^{-1}$).

\section{Large Extra Dimensions}

\begin{figure}
\mbox{\epsfig{file=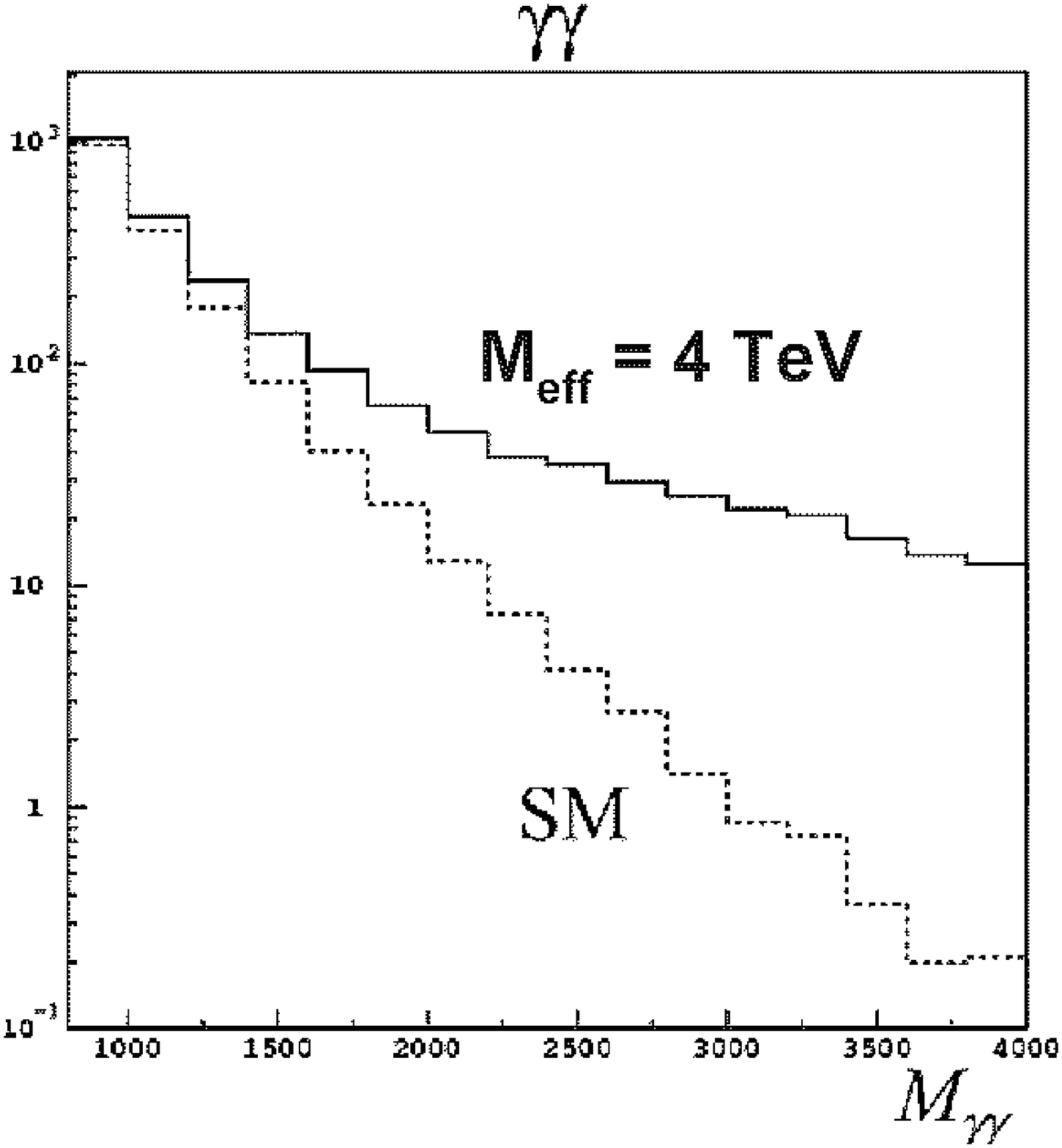,width=0.23\textwidth}}\hfill
\mbox{\epsfig{file=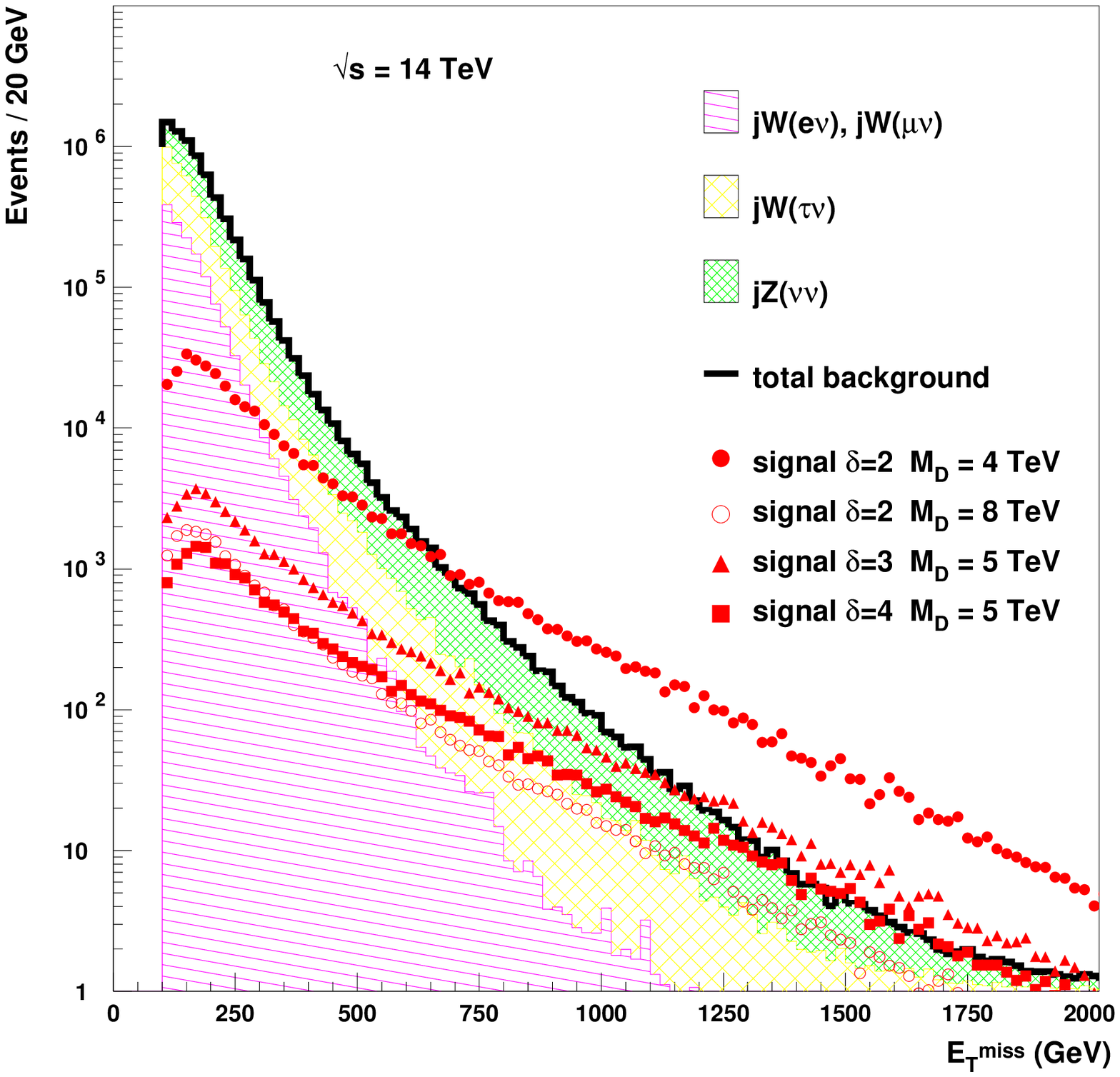,width=0.25\textwidth}}
\caption{{\bf Large ED.} 
Left plot~\cite{virtled}: virtual exchange of gravitons. The plot shows the deviation in 
Drell-Yan cross-section $pp\rightarrow\gamma\gamma$ (top curve) with respect to the SM expectation~\cite{virtled}.
Right plot~\cite{led}: direct production. Missing energy distribution 
(dots), shown here for various choices 
of the number of ED ($\delta$) and of the mass scale ($M_D$) and for SM 
 backgrounds 
(histograms).}
\label{fig:led}
\end{figure}

In this scenario, the SM fields are confined in our 4D world 
and only gravity propagates in the bulk.
The model is characterized by the number of extra dimensions $\delta$ 
and by the new fundamental scale $M_D$.
The graviton expands in 4D into a tower of Kaluza-Klein (KK) excitations 
which couple universally to all SM fields.
Even though this coupling is small ($1/M_{Pl}$), the large number of states and their small mass splitting lead to sizeable cross-sections at the 
LHC.

The virtual exchange of KK excitations of graviton 
can lead to deviations in 
Drell-Yan cross-sections and asymmetries in 
SM processes. 
The left plot of Fig.~\ref{fig:led} illustrates such deviations 
in the $\gamma\gamma$ invariant mass distribution~\cite{virtled}.
This kind of signatures is clear, very sensitive to new physics and could 
signal the existence of extra dimensions. 
However the underlying parameters of the model cannot be extracted in 
this case because the model is sensitive to unknown ultra-violet physics.

The second class of signatures is the direct production of KK 
excitations of graviton which will escape detection in 4D:
$q\bar{q}\rightarrow g G^{(k)}$, $gq\rightarrow q G^{(k)}$ and $gg\rightarrow g G^{(k)}$. 
In this case, the main signature to look for is some missing energy 
accompanied by a mono-jet (Fig.~\ref{fig:led}, right plot)~\cite{led}.
Within the allowed region for the effective theory ($\sqrt{\hat{s}}<M_D$), 
those processes can be reliably calculated and the parameters of the model 
can be constrained from the measurements.
Models with up to four extra dimensions could be probed at LHC. 
For 100 fb$^{-1}$, the maximum reach in $M_D$ is between 9.1 TeV ($\delta=2$) and 6.0 TeV ($\delta=4$), corresponding to a radius of compactification 
between 8 $\mu$m and 1 pm.
The two parameters of the model can in principle be extracted from the 
absolute cross-section of the processes or more definitely
 by collecting $\sim50$ fb$^{-1}$ of 
data at a different center-of-mass energy. 

\section{TeV$^{-1}$-sized Extra Dimensions}

\begin{figure}
\mbox{\epsfig{file=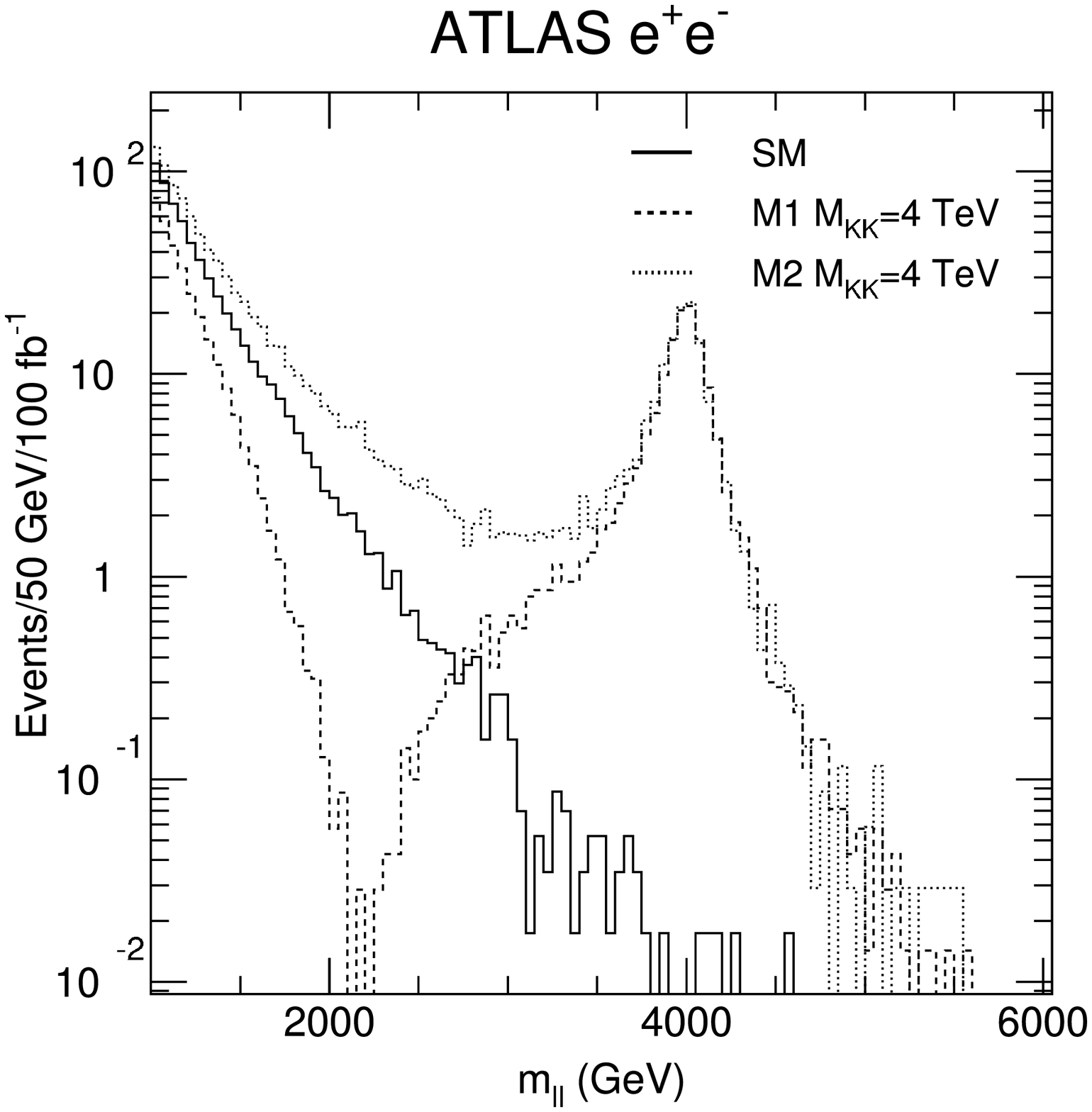,width=0.24\textwidth}}\hfill
\mbox{\epsfig{file=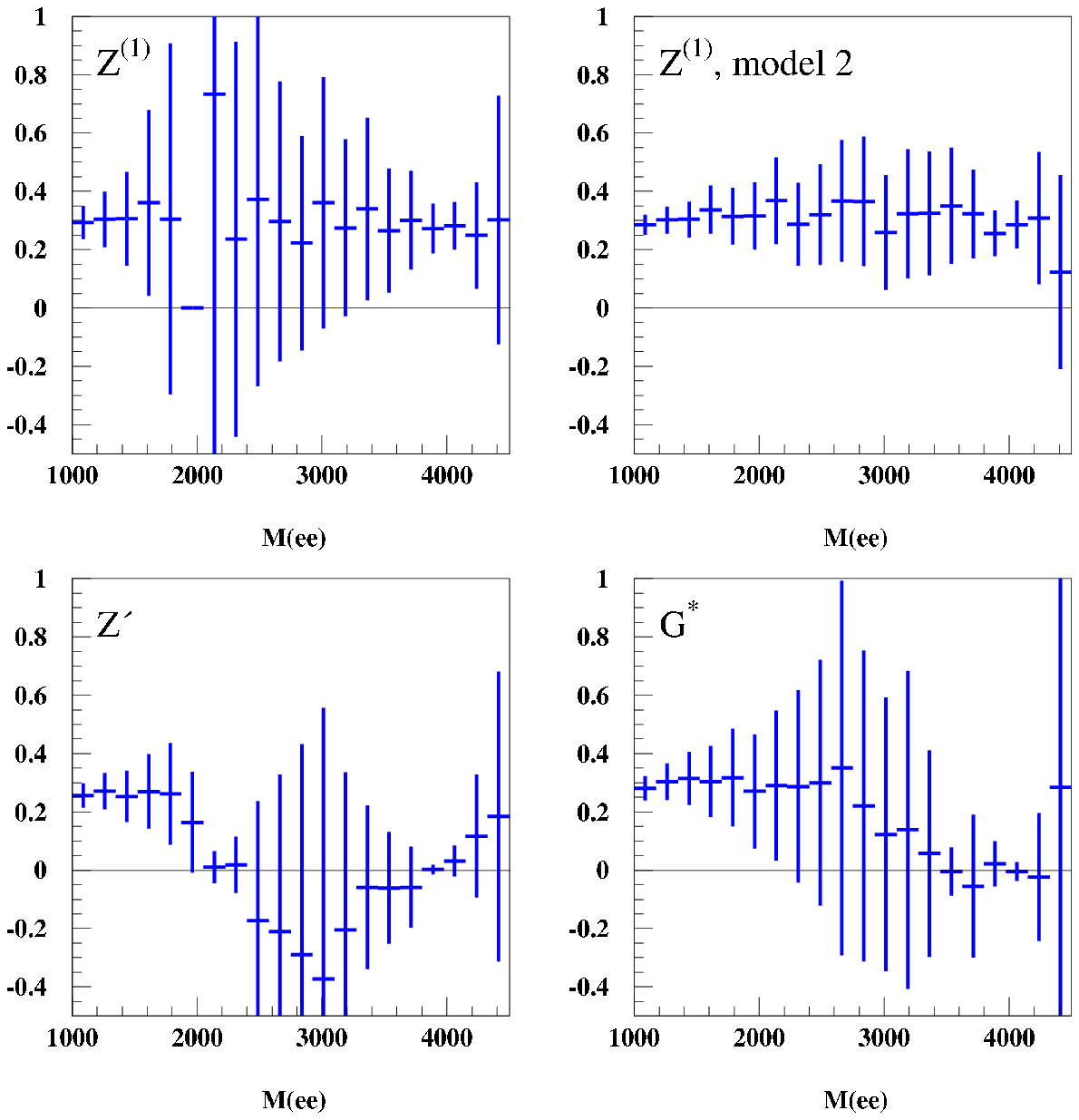,width=0.24\textwidth}}
\caption{ {\bf TeV$^{-1}$-sized ED.} 
Left plot~\cite{kkz}:
invariant mass distribution of $e^+e^-$ pairs for 
the SM (full line) and for two models where 
$Z^{(1)}/\gamma^{(1)}\rightarrow e^+e^-$
and $m_{Z^{(1)}}=4$ TeV.
Right plot~\cite{kkz}:
forward-backward asymmetries in the electron channel 
for different types of resonances, centered at $m=4$ TeV: 
two models for the $Z^{(1)}$ in $TeV^{-1}$ ED models (top), a $Z'$ 
(bottom left) and the graviton excitation in RS scenario.
}
\label{fig:kkz}
\end{figure}

In this case, the ED are small enough 
to allow the 
propagation of gauge bosons in the bulk without contradicting 
the existing electroweak measurements.
Only models with one small ED are considered here, with 
the fermions localized in the brane at different locations (M1 and 
M2 models).
The main feature of this model is the production of 
KK excitations of the $Z$ and $W$ bosons.
Their mass spectrum is defined by: 
$m_{Z^{(k)},W^{(k)}}^2 = m_{Z,W}^2 + k^2 M_C^2$ where $M_C$ is the 
compactification scale, known to be $\geq 4$ TeV 
from precision electroweak data.
Hence only the first excitations $Z^{(1)}$ and $W^{(1)}$ can be seen
at LHC.
In the electron 
channel~\cite{kkz} the experimental resolution is smaller than the natural width 
of the $Z^{(1)}$.
The expected signal~\footnote{NB: this plot is at parton level. 
The simulation study gives similar results.}
is shown on Fig.~\ref{fig:kkz}, left plot. 
The direct observation of a peak is possible if $M_C\leq 5.8$ TeV.
However, this reach can be improved drastically by using all the information with 
 a maximum 
likelihood, 
 in particular using the region before the peak 
(Fig.~\ref{fig:kkz}, left) with 
 either constructive or destructive 
interferences between the $Z^{(1)}/\gamma^{(1)}$ and $Z/\gamma$.
The sensitivity is thus increased to $M_C^{max}=9.5$ TeV and could even 
reach 13.5 TeV 
by combining the electron and muon channels for 300 fb$^{-1}$.

Furthermore, the spin-1 $Z^{(1)}$ signal can be distinguished from 
a spin-2 narrow graviton resonance (section~\ref{sec:rs}) using the angular distribution of 
its decay products.
Thanks to the contributions of the higher lying states,  
 the interference terms and to the additional $\sqrt{2}$ factor in its 
coupling to SM fermions, the $Z^{(1)}$ can also be distinguished 
from a $Z'$ with SM-like couplings:
 the distribution of the forward-backward asymmetry for the various cases are 
shown on Fig.~\ref{fig:kkz} for 4 TeV resonances.
The $Z^{(1)}$ hypothesis can be discriminated for masses up to about 5 TeV with 
an integrated luminosity of 300 fb$^{-1}$.

\section{Warped Extra Dimensions}
\label{sec:rs}

\begin{figure}
\mbox{\epsfig{file=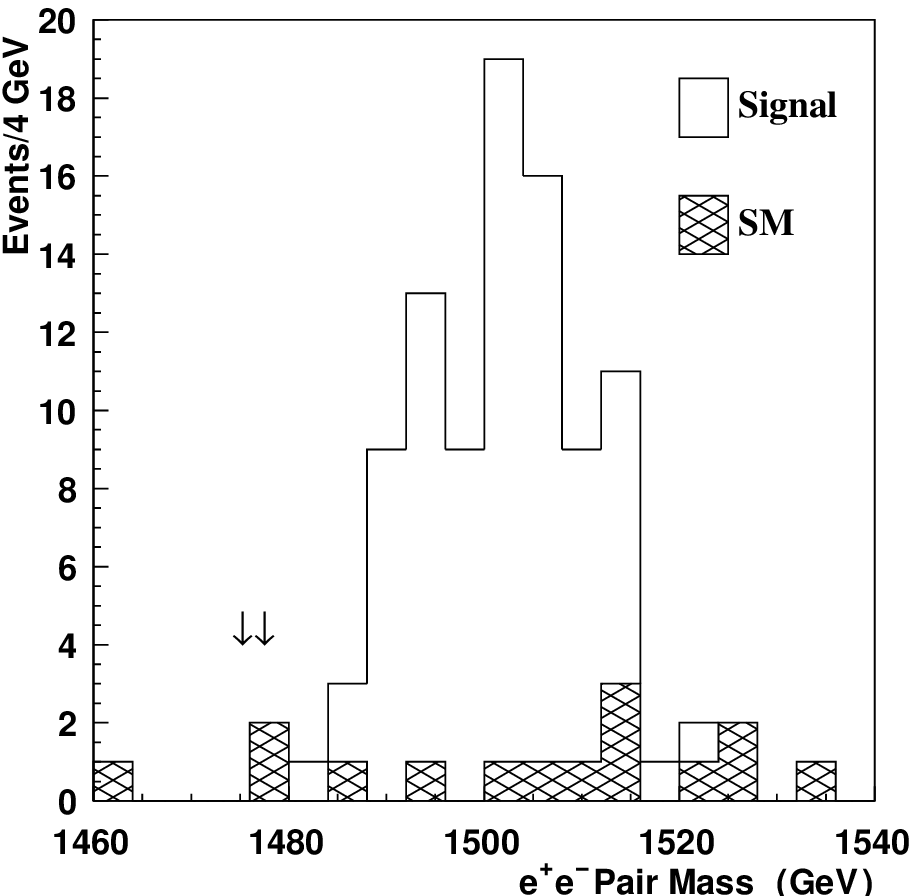,width=0.24\textwidth}}\hfill
\mbox{\epsfig{file=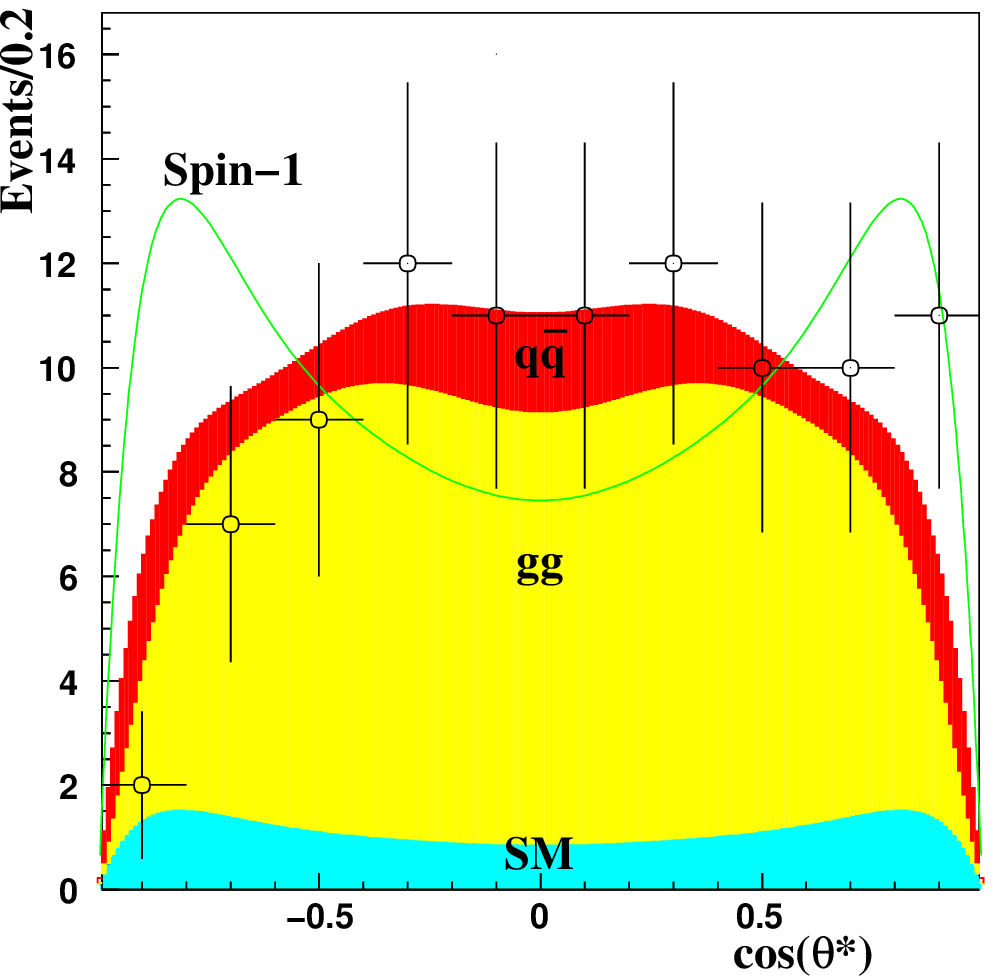,width=0.24\textwidth}}
\caption{ {\bf Warped ED.} 
Left plot~\cite{rs1}: $e^+e^-$ invariant mass distribution from 
graviton narrow resonance (open histogram) 
on top of SM background.
Right plot~\cite{rs1}:
angular distribution of $e^+e^-$ pairs for the 
graviton narrow resonance (open circles, yellow curve for the $gg$ 
dominant production and red curve for the $q\bar{q}$ production), 
for the SM (bottom blue curve) and the expected 
distribution for a spin 1 resonance (green line).
}
\label{fig:rs1}
\end{figure}

\begin{figure}
\mbox{\epsfig{file=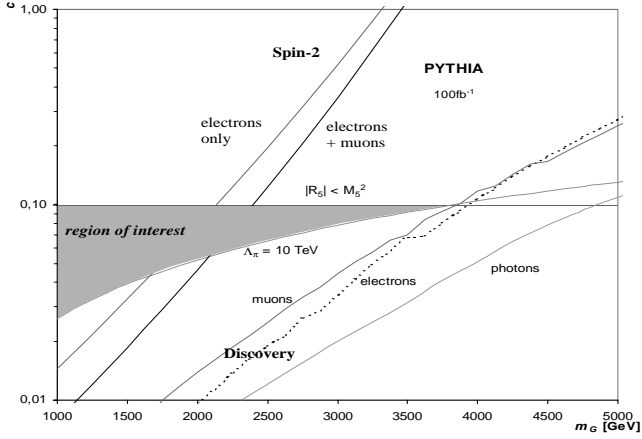,height=6.0cm,width=0.48\textwidth}}
\caption{ {\bf Warped ED.}
Exclusion limits for discovery of a RS graviton resonance as a function 
of the mass and of the curvature scale $c=k/M_{Pl}$. 
See text for explanations.
\cite{rs2}.}
\label{fig:rs2}
\end{figure}

We consider here one of the Randall-Sundrum models of warped ED, 
where gravity propagates in a 5D 
bulk limited by two branes. 
The SM fields are confined in the first brane.
The metric includes an exponential warp factor which curves the space and 
connects the 
$M_{EW}$ scale in the SM brane to the Planck scale in the other brane, 
hence reformulating the hierarchy problem.
The phenomenology is defined by two parameters: the scale of physical processes 
in the SM brane, $\Lambda_\pi \sim 1$ TeV, and the curvature scale $c=k/M_{Pl}$.
The main feature of this model is the production of KK graviton narrow resonances,
whose masses are given by $m_n = x_n \Lambda_\pi c$ where $x_n$ are the roots of 
the $J_1$ Bessel function.
Here again only the first excitation is likely to be seen at LHC, the other modes 
being suppressed by the falling parton distribution functions.
Some of the best channels are $G^{(1)}\rightarrow e^+e^-$ and 
$G^{(1)}\rightarrow \gamma\gamma$, thanks to the energy and angular resolutions 
of the LHC detectors.
The signal expected with 100 fb$^{-1}$ in the channel $e^+e^-$ is shown on 
Fig.~\ref{fig:rs1}, left plot, for a pessimistic hypothesis of $c=0.01$.
In this case, masses up to 2 TeV can be probed~\cite{rs1}. 
The reach goes up to about 4 TeV for $c=0.1$.
The sensitivity is summarized on Fig.~\ref{fig:rs2} in the $(m_{G^{(1)}},c)$ plane~\cite{rs2}: 
if no signal is found, the area left to the 
curves labeled "Discovery" can be excluded at 95\% CL. 
The interesting region in this plane is limited by the constraint 
$\Lambda_\pi \leq 10$ TeV (otherwise the model would 
 no longer be interesting for solving the hierarchy problem) and by the 
$c<0.1$ limit (low-energy consistency).
After one year of data-taking the LHC covers completely this region.

The spin-2 nature of $G^{(1)}$ can be measured as well as shown on 
Fig.~\ref{fig:rs1}, right plot.
It is worth noting that the acceptance at large pseudo-rapidities ($1.5<|\eta|<2.5$) 
is essential for the spin discrimination.
The curves labelled "Spin-2" on Fig.~\ref{fig:rs2} show that for graviton 
masses up to 2.3 TeV ($c=0.1$), there is a 90\% chance that the spin-2 nature of 
the graviton can be determined with a 95\% CL.
It has also been shown~\cite{rs1} that this resonance can be seen 
in many other channels ($\mu\mu$,$\gamma\gamma$,$jj$,$b\bar{b}$,
$t\bar{t}$,$WW$,$ZZ$), hence allowing to check the universality of its couplings;
and that the size $R$ of the ED could also be estimated with a 10\% accuracy from the 
mass and cross-section measurements.

\begin{figure}
\mbox{\epsfig{file=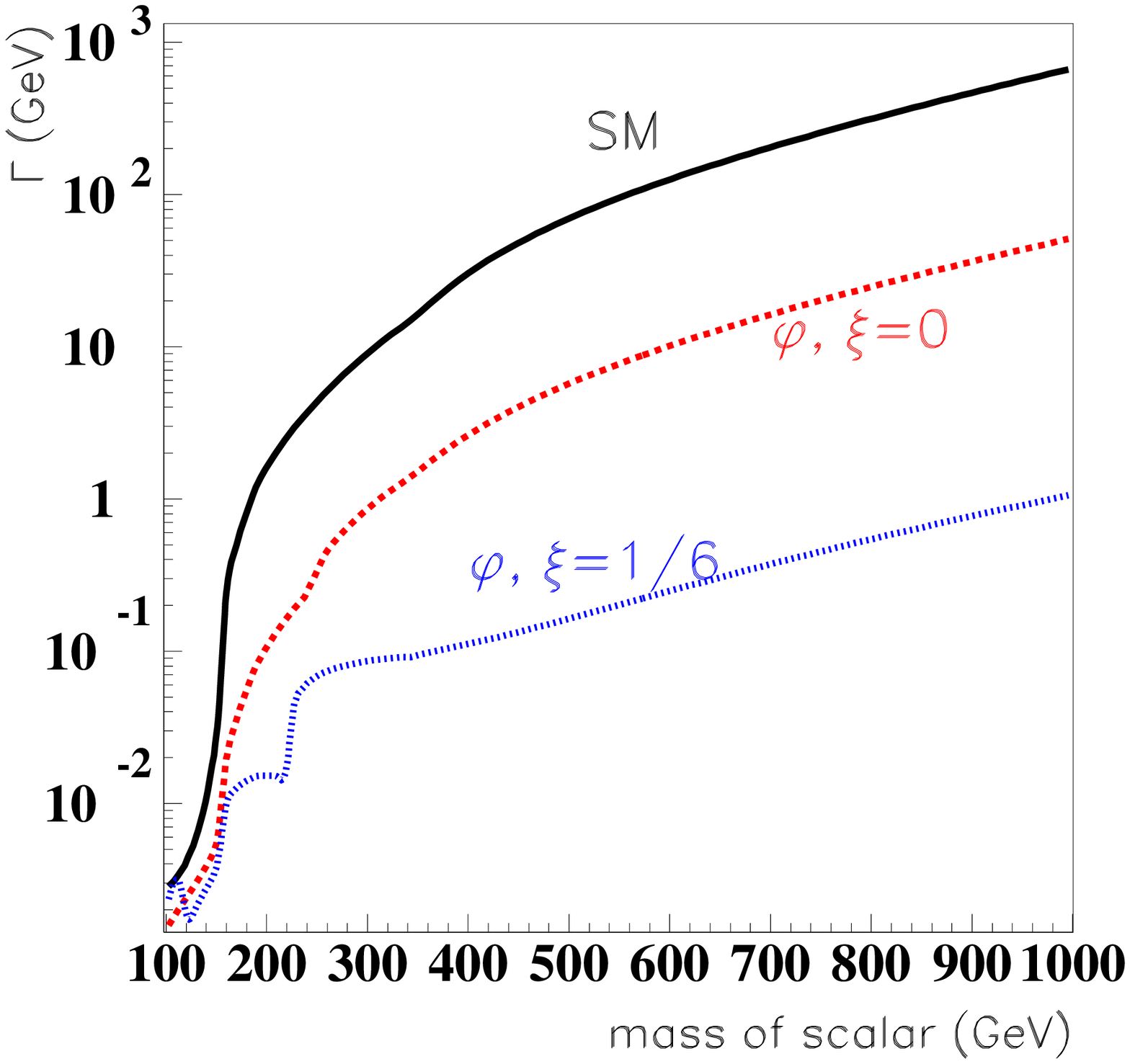,height=4.8cm,width=0.22\textwidth}}\hfill
\mbox{\epsfig{file=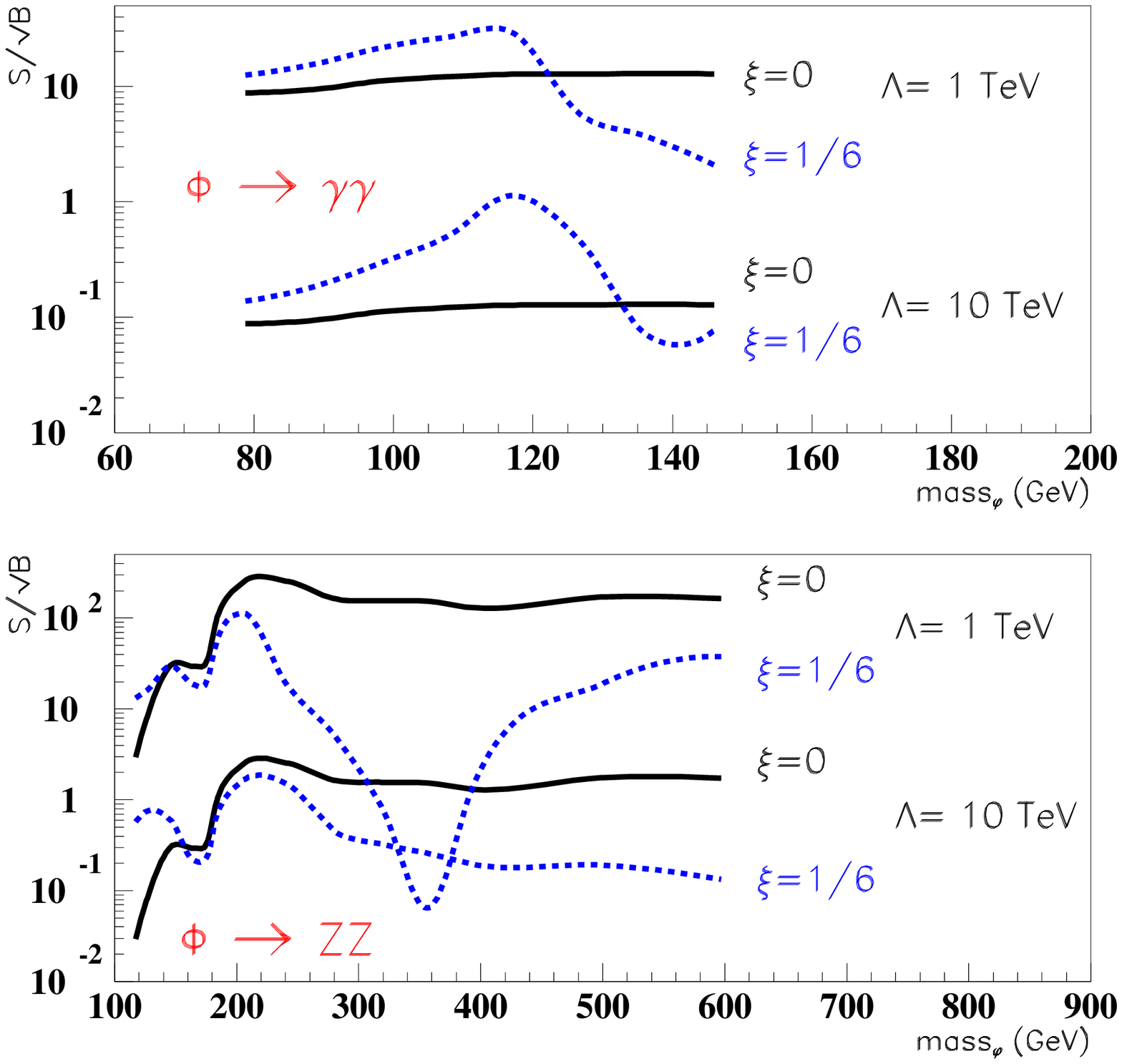,width=0.27\textwidth}}
\caption{ {\bf Warped ED.}
Left plot~\cite{radion}: width of the radion compared to the one of the SM Higgs.
Right plot~\cite{radion}:
significances of two radion decay channels as a function 
of the radion mass. For $\xi=0$ there is no radion-Higgs mixing 
while for $\xi=1/6$ they are heavily mixed.
}
\label{fig:phi}
\end{figure}

Another intriguing signature of this model is the existence of a new massive 
scalar, the radion, allowing to stabilize the spacing between the two branes 
at the distance required for solving the hierarchy problem ($kR\sim12$).
The radion is very similar to the SM Higgs, and can actually mix with it. 
Its width is smaller though, as shown on Fig.~\ref{fig:phi}, left plot; and its 
partial widths can be different, with in particular an enhanced coupling 
to gluons.
There are three additional parameters for the radion: its mass, the vacuum 
expectation value $\Lambda_\phi$ and the radion-Higgs mixing parameter $\xi$.
The results for the SM Higgs have been reinterpreted for the radion case by 
folding in the new branching ratios and the appropriate detector resolutions~\cite{radion}.
Fig.~\ref{fig:phi} (right plot) shows the expected significances for a 
radion signal in the $\gamma\gamma$ and $ZZ^{(*)}$ channels.
Discovery is possible over the whole mass range if $\Lambda_\phi\sim$ TeV.
If its mass permits, the radion can also decay into a pair of Higgs scalars.
If $m_\phi=300$ GeV, only 2(4) fb$^{-1}$ are needed for a $5\sigma$ discovery 
in the very clean channel $\phi\rightarrow hh \rightarrow \gamma\gamma b\bar{b}$ for $\xi=0 (1/6)$.
With 30 fb$^{-1}$ collected at low luminosity (to maintain excellent $b$-tagging), 
scales up to $\Lambda_\phi=2.2$ TeV can be probed in this channel, and up to 1.0 TeV in the 
$\phi\rightarrow hh \rightarrow b\bar{b} \tau^+\tau^-$ for $m_\phi=600$ GeV.
Discrimination between the radion and a Higgs scalar will require very precise 
measurements of their couplings.

\section{Other models and signatures}

The existence of ED could also be probed with the di-jet 
cross-section~\cite{laforge} or via the polarization of 
$\tau$ leptons in some 
 models with two Higgs doublets and a singlet 
neutrino in the bulk~\cite{ketevi}.
Finally, the production of black holes at LHC is another striking 
signature of ED~\cite{greg}.

\section{Conclusion}

The LHC will be able to probe the relevant region of the 
parameter space for most of the models with ED
 studied so far.
Moreover, in most cases it will be possible to discriminate such 
signals from other new physics scenarios and infer some 
information about the underlying model.

\section*{Acknowledgements}

The studies presented here have been performed within the ATLAS 
and CMS collaborations and as such have made use of tools which 
are the results of collaboration-wide efforts.
I would like to thank in particular G.~Azuelos, F.~Gianotti, 
I.~Hinchliffe,
L.~Pape, L.~Poggioli, S.~Shmatov, P.~Traczyk and 
G.~Wrochna.

\end{document}